\documentclass[a4paper,10pt,conference]{IEEEtran}
\IEEEoverridecommandlockouts
\usepackage{cite}
\usepackage{amsmath,amssymb,amsfonts}
\usepackage{graphicx}
\usepackage{tabularx}
\usepackage{textcomp}
\usepackage{multirow}
\usepackage{changepage}
\usepackage{booktabs}
\usepackage{xcolor}
\usepackage{subcaption}

\def\BibTeX{{\rm B\kern-.05em{\sc i\kern-.025em b}\kern-.08em
    T\kern-.1667em\lower.7ex\hbox{E}\kern-.125emX}}
\begin{document}

\title{Security Evaluation of Compressible and Learnable Image Encryption against Jigsaw Puzzle Solver Attacks}

\author{\IEEEauthorblockN{1\textsuperscript{st} Tatsuya Chuman}
\IEEEauthorblockA{\textit{Tokyo Metropolitan University} \\
Tokyo, Japan \\
chuman-tatsuya1@ed.tmu.ac.jp}
\and

\IEEEauthorblockN{2\textsuperscript{st} Nobutaka Ono}
\IEEEauthorblockA{\textit{Tokyo Metropolitan University} \\
Tokyo, Japan \\
onono@tmu.ac.jp}
\and
\IEEEauthorblockN{3\textsuperscript{nd} Hitoshi Kiya}
\IEEEauthorblockA{\textit{Tokyo Metropolitan University} \\
Tokyo, Japan \\
kiya@tmu.ac.jp}
}

\maketitle
\setlength{\abovedisplayskip}{4pt}
\setlength{\belowdisplayskip}{4pt}
\setlength\abovecaptionskip{1mm}
\setlength\textfloatsep{1mm}
\setlength\floatsep{0pt}
\setlength\parskip{.2mm}

\begin{abstract}
Several learnable image encryption schemes have been developed for privacy-preserving image classification. This paper focuses on the security of block-based image encryption methods that are learnable and JPEG-friendly. Permuting divided blocks in an image is known to enhance robustness against ciphertext-only attacks (COAs), but recently jigsaw puzzle solver attacks have been demonstrated to be able to restore visual information on the encrypted images. In contrast, it has never been confirmed whether encrypted images including noise caused by JPEG-compression are robust. Accordingly, the aim of this paper is to evaluate the security of compressible and learnable encrypted images against jigsaw puzzle solver attacks. In experiments, the security evaluation was carried out on the CIFAR-10 and STL-10 datasets under JPEG-compression. 
\end{abstract}

\begin{IEEEkeywords}
Image Encryption, Vision Transformer
\end{IEEEkeywords}

\section{Introduction}
Nowadays, the remarkable development of deep neural networks (DNNs) makes it possible to solve complex tasks for many applications, including privacy-sensitive security-critical ones such as facial recognition and medical image analysis.
Although deep learning for image classification on a cloud platform is an effective choice for a user owing to its cost and ease of use, an image including privacy information tends to be processed on the premises due to the risk of data leakage.
Numerous image encryption schemes have been proposed for privacy-preserving image classification to protect visual information on images \cite{warit2018ICME,tanaka_AAAI, kiya2022seg, kiya_IEICE}, but several ciphertext-only attacks (COAs) including DNN-based ones were shown to restore visual information on encrypted images \cite{kiya2022overview}.
Therefore, encryption schemes that are robust against various attacks are essential for privacy-preserving image classification.
In addition to being robust against attacks, ensuring high image classification accuracy is essential for encryption schemes.
\par
In contrast, encryption methods for  application to the vision transformer (ViT)  \cite{dosovitskiy2021an} have been proposed for privacy-preserving deep learning \cite{AprilPyone_compress,AprilPyone_IEEEtrans,qi_vit_2022,2023_Imaizumi}.
These schemes allow us to perform deep learning with visually protected images while maintaining high performance that ViT has.
Furthermore, the compressible and learnable image encryption scheme was proposed for the purpose of reducing the amount of data \cite{2023_Imaizumi}.
\par
The jigsaw puzzle solver attack succeeded in restoring visual information of encrypted images for being applied to ViT \cite{2023_chuman_mdpi}.
On the other hand, it was confirmed that noise caused by JPEG-compression makes jigsaw puzzle solver attacks more difficult to restore visual information \cite{2006_Arnia, Chuman_IEEEtrans}.
In contrast, it has never been confirmed whether the encrypted images \cite{2023_Imaizumi} including noise caused by JPEG-compression are robust.
The aim of this paper is to evaluate the security of the compressible and learnable image encryption \cite{2023_Imaizumi} against the jigsaw puzzle solver attack.

\begin{figure}[t]
	\centering
	\includegraphics[width =\linewidth]{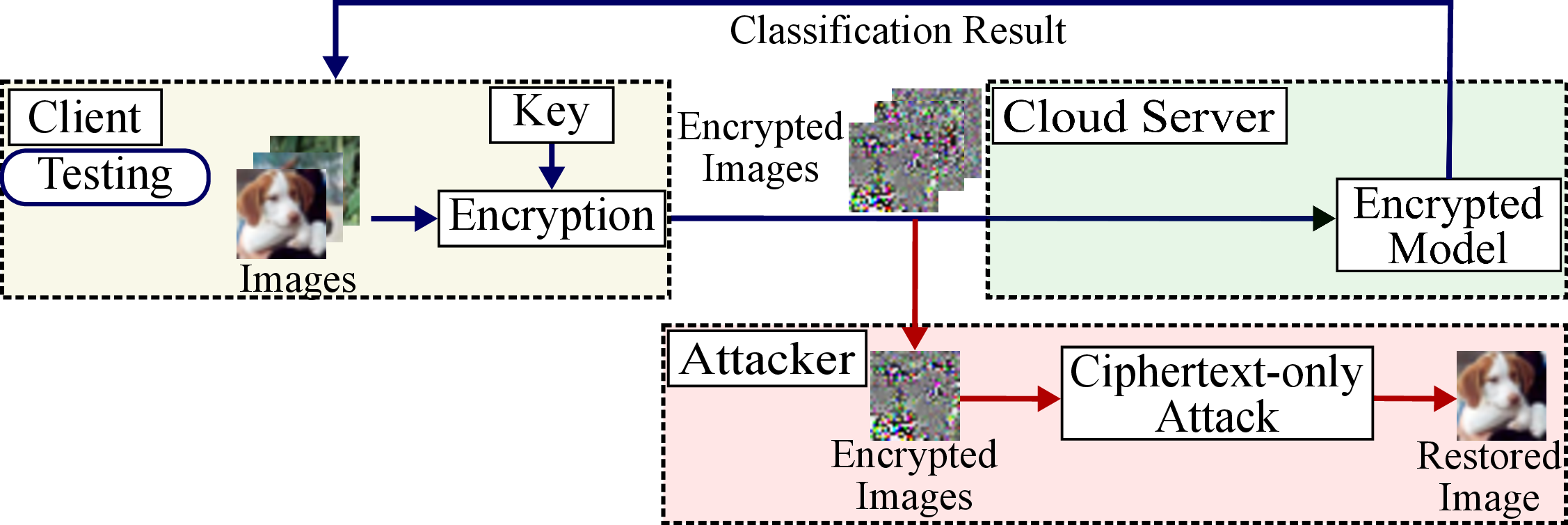}
	\caption{Scenario of restoring visual information of encrypted images by using ciphertext-only attacks}
	\label{fig:ppdn}
\end{figure}
\section{Framework of Security Evaluation}

\subsection{Overview}
Fig. \ref{fig:ppdn} shows the scenario of this paper.
A client sends an encrypted image to a cloud server, which has an encrypted model for image classification, to get a classification result.
However, there is a risk that an adversary could attempt to restore visual information on the encrypted image.
In this paper, we assume that the attacker knows access to encrypted images and the encryption algorithm but does not possess the secret key, which was used for encryption.

\subsection{Compressive and Learnable Image Encryption}
\label{sec:enc}
Several image encryption schemes for application to ViT were developed \cite{AprilPyone_IEEEtrans,qi_vit_2022,2023_Imaizumi}.
It has been known that ViT, a model for image classification based on the transformer architecture, is carried out by dividing an image into a grid of square patches\cite{dosovitskiy2021an}.
For example, images with 96$\times$96 pixels in the STL-10 dataset are resized to 224$\times$224 or 384$\times$384 pixels, and then divided into 16$\times$16 patch to fit the same patch size of pre-trained model such as ViT-B/16 and ViT-L/16.
This paper focuses on the security of the block-based perceptual image encryption method that is learnable and JPEG-friendly \cite{2023_Imaizumi}.
\par
\begin{figure}[t]
	\captionsetup[subfigure]{justification=centering}
	\centering
	\subfloat[Original image\newline($X \times Y$ = $224
	\times 224$)] {\includegraphics[clip, width=3cm]{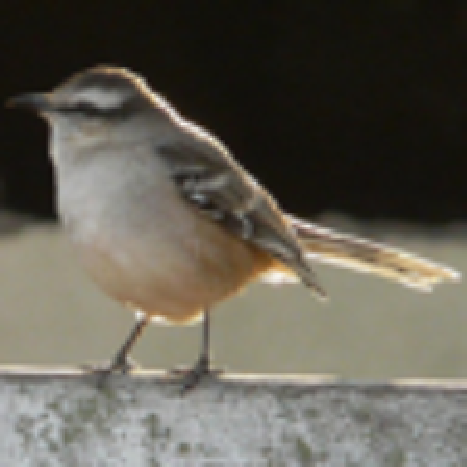}
	\label{subfig:oriimg}}
	\hfil
	\subfloat[Encrypted image \cite{2023_Imaizumi} ($M=16$)]
	{\includegraphics[clip, width=3cm]{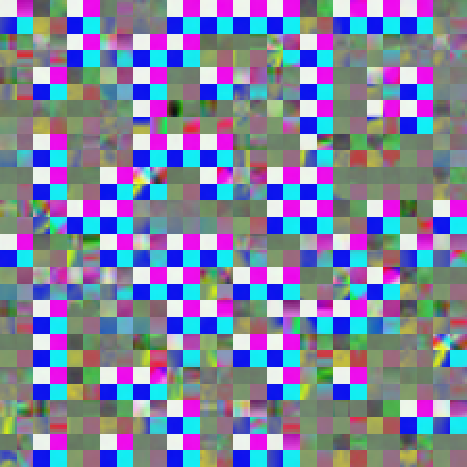}
	\label{subfig:encimg}}
	\caption{Example of encrypted image}
	\label{fig:eximg}
\end{figure}

The procedure of the compressive and learnable image encryption \cite{2023_Imaizumi} for a 24-bit RGB color image is described below.
\begin{itemize}
	\setlength{\leftskip}{0.5cm}
	\item[Step 1:]Divide an image with $X \times Y$ pixels into non-overlapped blocks with $M \times M$ pixels. In this study, $M=16$ is selected as used in \cite{2023_Imaizumi}.
	\item[Step 2:] Permute randomly the divided blocks using a random integer generated by a secret key $K_1$.
	\item[Step 3:] Split each divided block to generate sub-blocks with $\frac{M}{2} \times \frac{M}{2}$ pixels.
	\item[Step 4:] Permute randomly the sub-blocks within the block using a random integer generated by a secret key $K_2$. 
	\item[Step 5:] Rotate and invert randomly each sub-block by using a random integer generated by a key $K_3$.
	\item[Step 6:] Apply negative-positive transformation to each sub-block by using a random binary integer generated by a key $K_4$. In this step, a pixel value $p$ in sub-block is transformed to $p'$ by
	\begin{equation}
		\label{negaposi}
	p'=
	\left\{
	\begin{array}{ll}
	p & (r(i)=0) \\
	p \oplus (2^8-1) & (r(i)=1)
	\end{array} ,
	\right.
	\end{equation}
	where $r(i)$ is a random binary integer generated by the secret key. In this paper, the value of occurrence probability $P(r(i)) = 0.5$ is used to invert bits randomly.
	\item[Step 7:] Shuffle three color components commonly in each sub-block by using an integer randomly selected from six integers by a key $K_5$.
	\item[Step 8:] Integrate all sub-blocks to generate an encrypted image.
\end{itemize}
In this paper, although the keys $K_1$ and $K_2$, are commonly used among all color components, $K_3$ and $K_4$ are independently used among color components.
An example of encrypted images is illustrated in Fig. \ref{fig:eximg} (\subref{subfig:encimg}); Fig. \ref{fig:eximg} (\subref{subfig:oriimg}) shows the original one. 


\section{Jigsaw Puzzle Solver-based Attack}
Although permuting divided blocks in an image enhances robustness against COAs, several jigsaw puzzle solver attacks are known to be effective for restoring visual information \cite{Chuman_IEEEtrans, 2023_chuman_mdpi}.
Since the operation of encryption in the block-based image encryption scheme for application to ViT is performed using a common secret key for all sub-blocks, the jigsaw puzzle solver-based attack using edge information in each sub-block was proposed \cite{2023_chuman_mdpi}.
On the other hand, it has never been confirmed whether images including noise caused with JPEG-compression is robust enough against the jigsaw puzzle solver-based attack.
Therefore, in this paper, we evaluate the security of the compressible and learnable image encryption \cite{2023_Imaizumi} against the jigsaw puzzle solver-based attack.
\par
\begin{figure}[t]
	\centering
	\includegraphics[width =\linewidth]{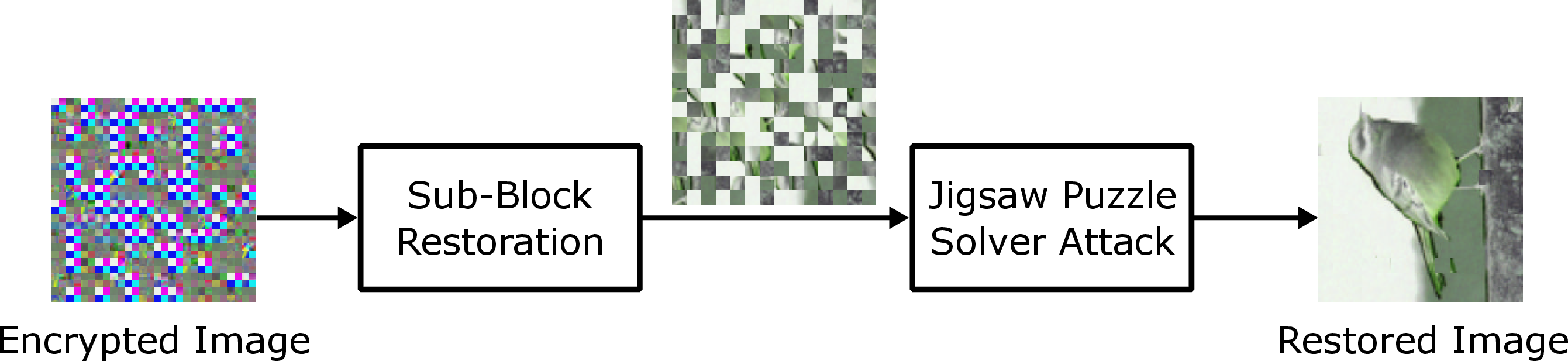}
	\caption{Jigsaw puzzle solver-based attack}
	\label{fig:jpsb}
\end{figure}

The jigsaw puzzle solver-based attack consists of two steps as illustrated in Fig. \ref{fig:jpsb}: sub-block restoration and jigsaw puzzle solver attack.
The purpose of the sub-block restoration is to solve the encryption in sub-blocks, whereas the jigsaw puzzle solver attack aims to assemble permuted blocks.
\par
Since the process of sub-block restoration depends on the encryption algorithm, the sub-block restoration in \cite{2023_chuman_mdpi} was extended to apply the compressible and learnable image encryption\cite{2023_Imaizumi}.
Namely, the encrypted image including RGB shuffled, negative-positive transformed, rotated, inverted and permuted sub-blocks as described in Sec.\ref{sec:enc} is solved by using the sub-block restoration for the compressible and learnable image encryption \cite{2023_Imaizumi}.
After solving the encryption in sub-block, the permuted blocks are assembled by using the jigsaw puzzle solver \cite{Sholomon_2016_GPEM}.

\begin{figure*}[t]
	\captionsetup[subfigure]{justification=centering}
	\begin{minipage}[b]{\linewidth}
	\centering
	\includegraphics[clip, width=9cm]{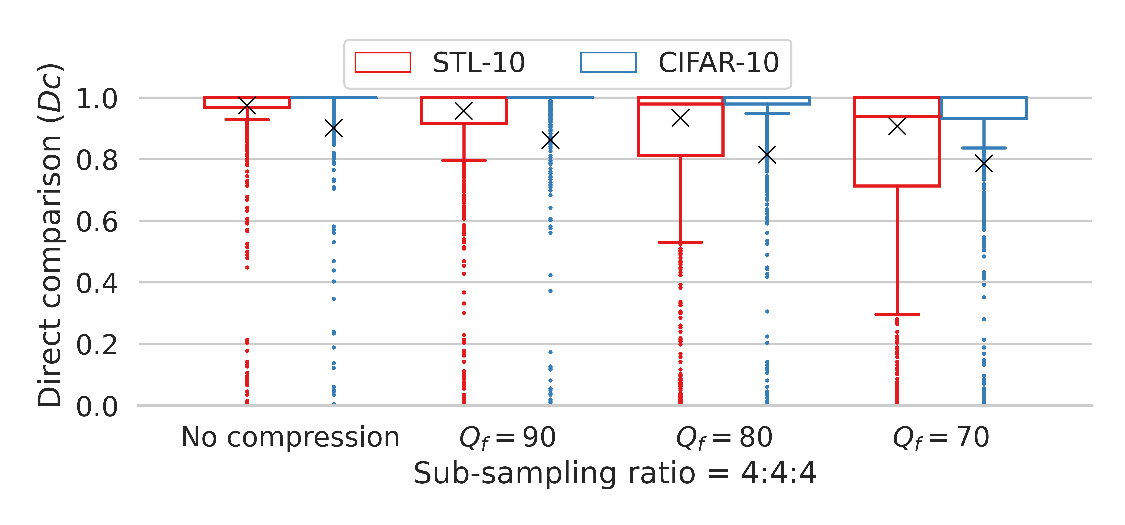}
	\includegraphics[clip, width=9cm]{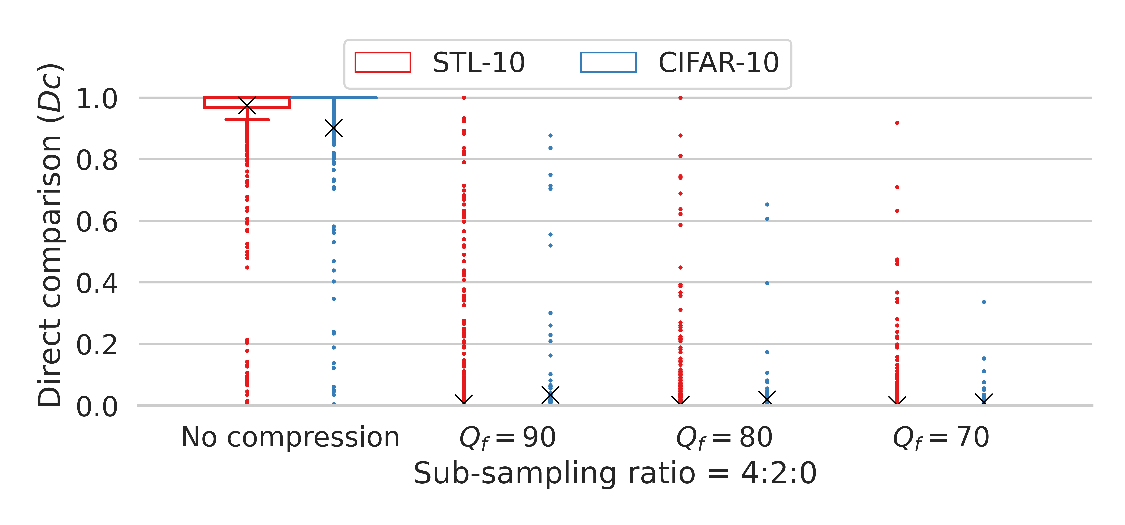}
	\label{subfig:dc}
	\vspace{-0.5cm}
	\subcaption{Direct comparison ($D_{c}$)}
	\end{minipage}
	\\
	\begin{minipage}[b]{\linewidth}
	\includegraphics[clip, width=9cm]{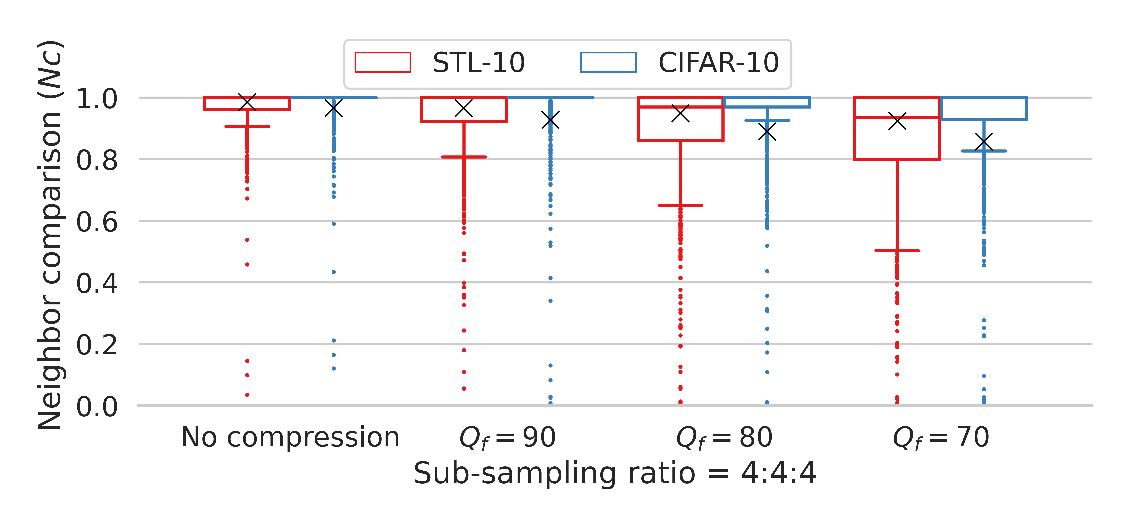}
	\includegraphics[clip, width=9cm]{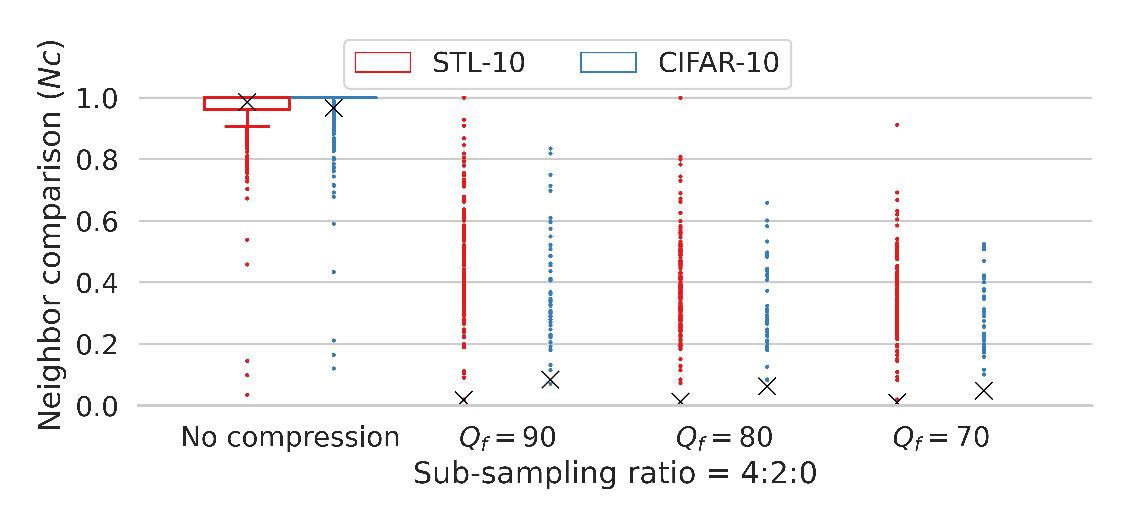}
	\label{subfig:nc}
	\vspace{-0.5cm}
	\subcaption{Neighbor comparison ($N_{c}$)}
	\end{minipage}
	\\
	\begin{minipage}[b]{\linewidth}
	\includegraphics[clip, width=9cm]{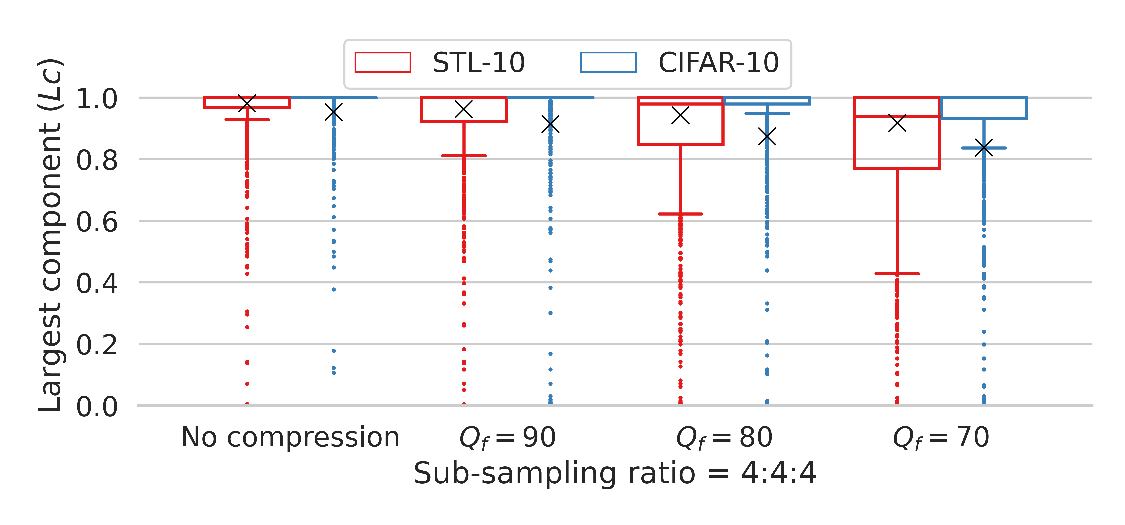}
	\includegraphics[clip, width=9cm]{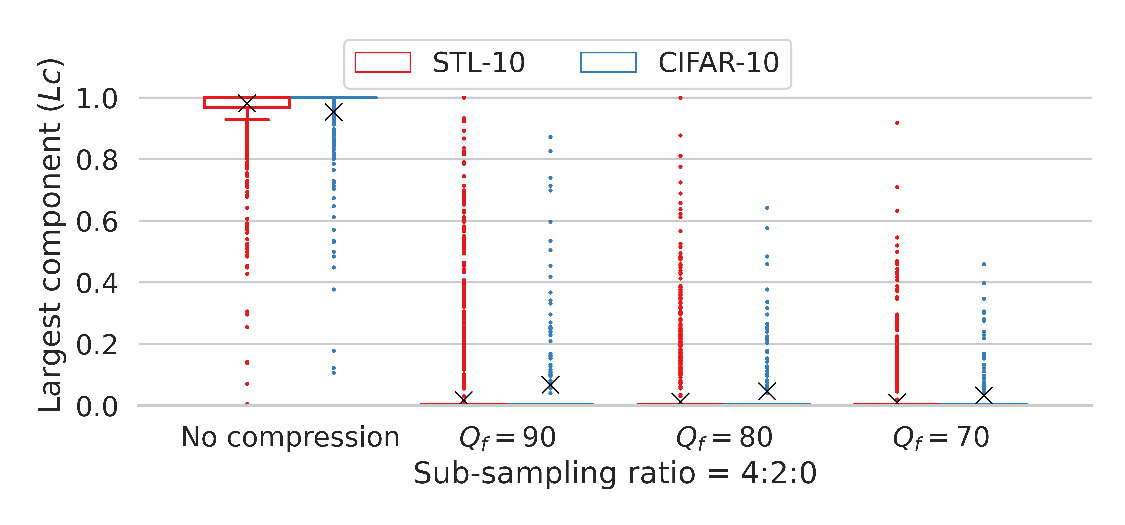}
	\label{subfig:lc}
	\vspace{-0.5cm}
	\subcaption{Largest component ($L_{c}$)}
	\end{minipage}
	\caption{Average direct comparison ($D_{c}$), largest component ($L_{c}$) and neighbor comparison ($N_{c}$) values of images reconstructed by using jigsaw puzzle solver-based attack. Boxes span from first to third quartile, referred to as $Q_{1}$ and $Q_{3}$, and whiskers show maximum and minimum values in range of [$Q_{1}$ - 1.5($Q_{3}$ - $Q_{1}$), $Q_{3}$ + 1.5($Q_{3}$ - $Q_{1}$)]. Band and cross indicate median and average values, respectively. Dots represent outliers.}
	\label{fig:dclc}
\end{figure*}

\newcommand{\oristone}{\includegraphics[width=4.5em]{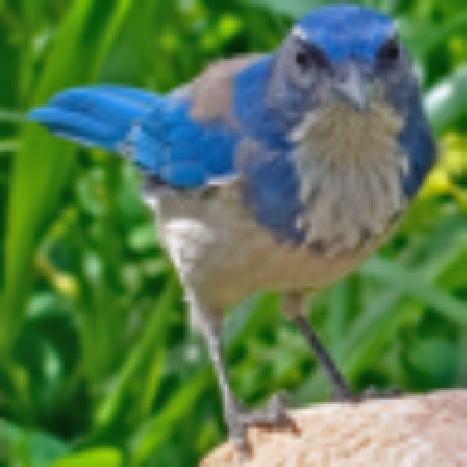}}
\newcommand{\oristtwo}{\includegraphics[width=4.5em]{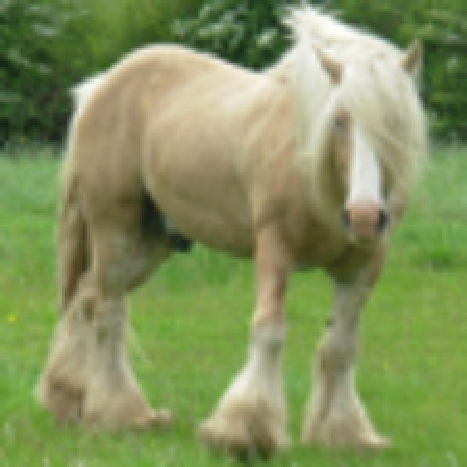}}
\newcommand{\oristthree}{\includegraphics[width=4.5em]{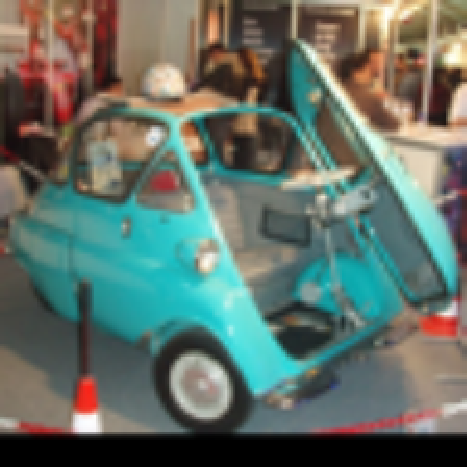}}
\newcommand{\oristfour}{\includegraphics[width=4.5em]{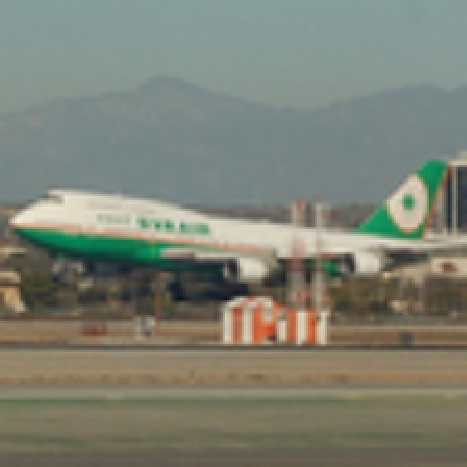}}

\newcommand{\encstone}{\includegraphics[width=4.5em]{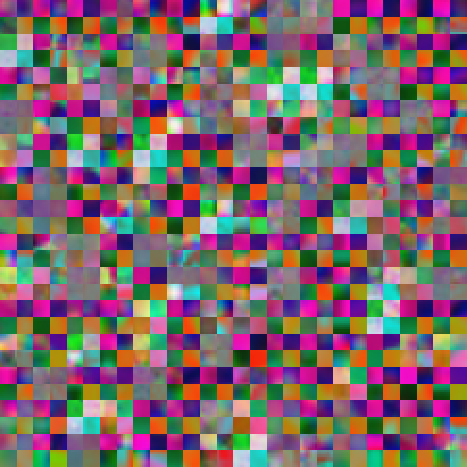}}
\newcommand{\encsttwo}{\includegraphics[width=4.5em]{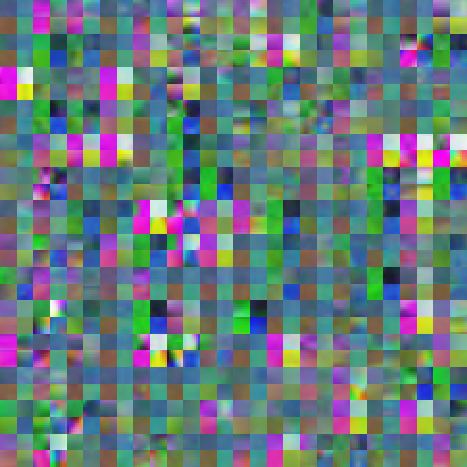}}
\newcommand{\encstthree}{\includegraphics[width=4.5em]{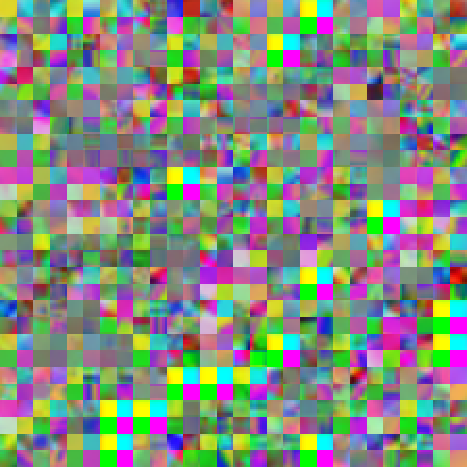}}
\newcommand{\encstfour}{\includegraphics[width=4.5em]{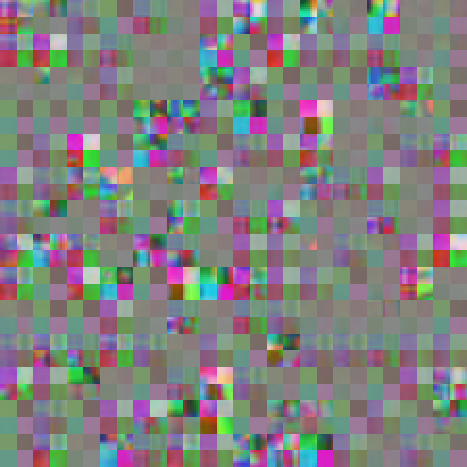}}

\newcommand{\resonestone}{\includegraphics[width=4.5em]{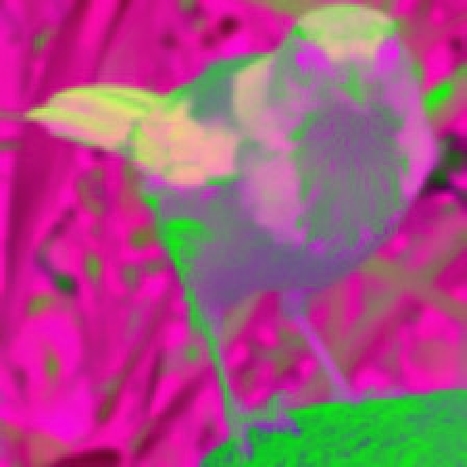}}
\newcommand{\resonesttwo}{\includegraphics[width=4.5em]{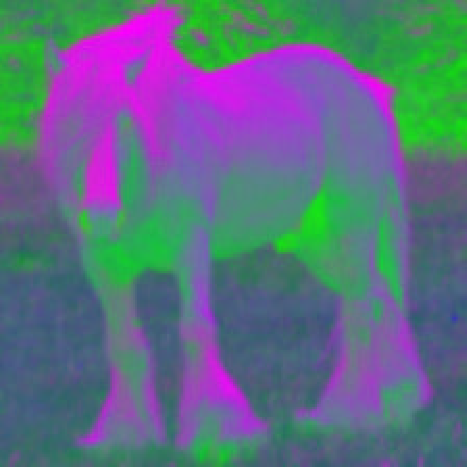}}
\newcommand{\resonestthree}{\includegraphics[width=4.5em]{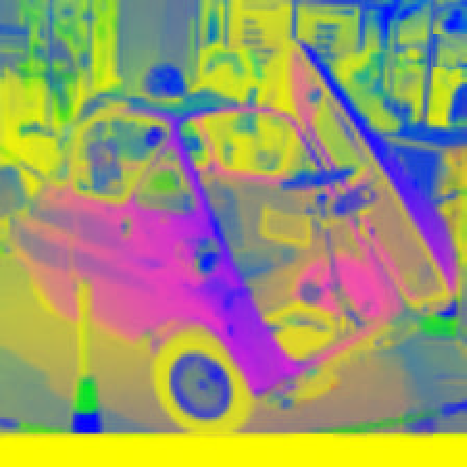}}
\newcommand{\resonestfour}{\includegraphics[width=4.5em]{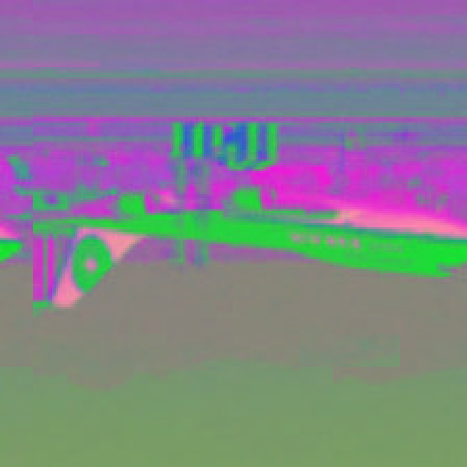}}

\newcommand{\restwostone}{\includegraphics[width=4.5em]{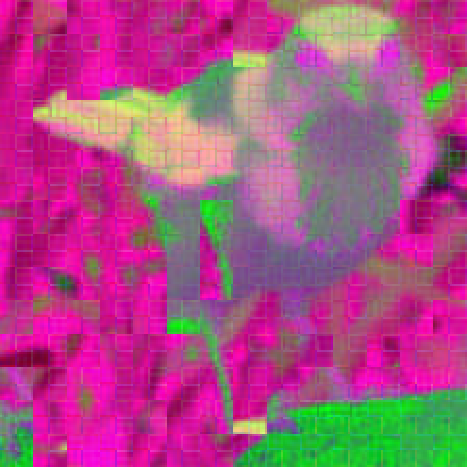}}
\newcommand{\restwosttwo}{\includegraphics[width=4.5em]{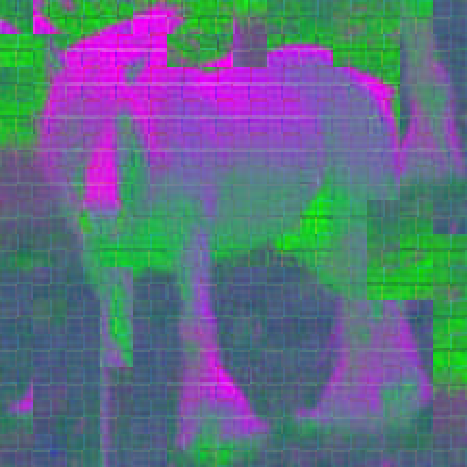}}
\newcommand{\restwostthree}{\includegraphics[width=4.5em]{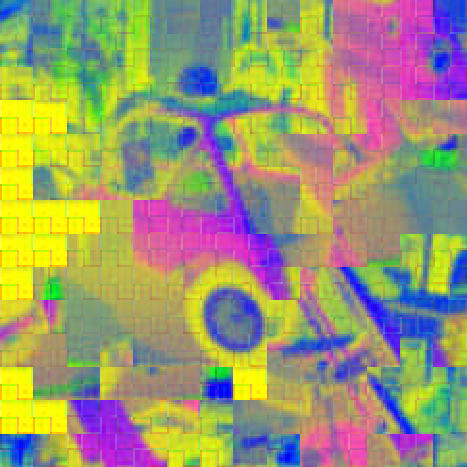}}
\newcommand{\restwostfour}{\includegraphics[width=4.5em]{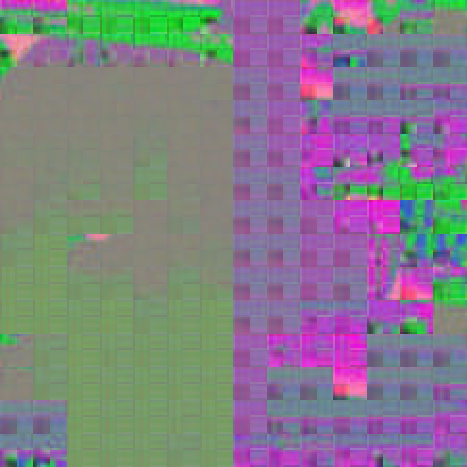}}

\newcommand{\coristone}{\includegraphics[width=4.5em]{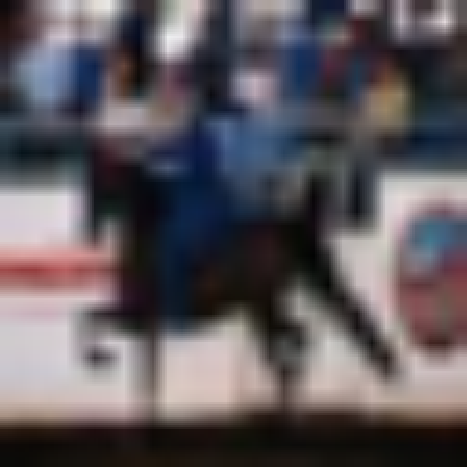}}
\newcommand{\coristtwo}{\includegraphics[width=4.5em]{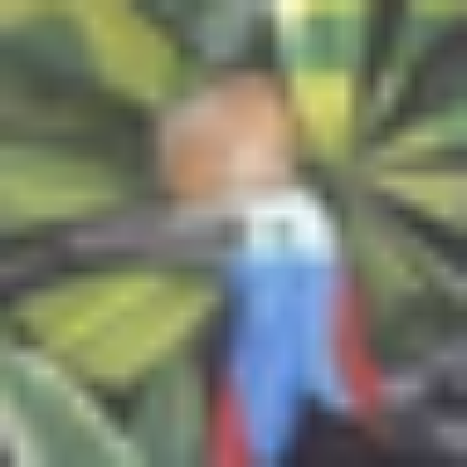}}
\newcommand{\coristthree}{\includegraphics[width=4.5em]{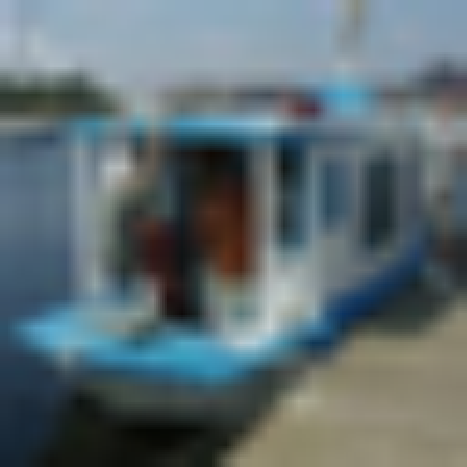}}
\newcommand{\coristfour}{\includegraphics[width=4.5em]{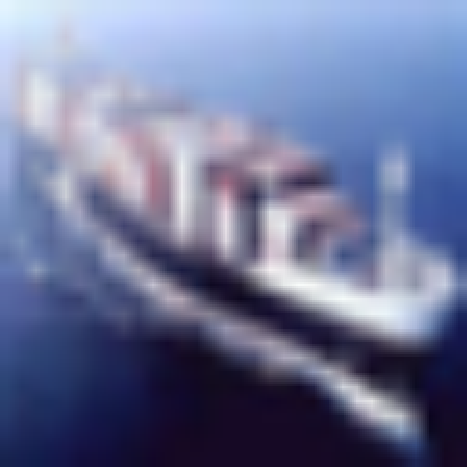}}

\newcommand{\cencstone}{\includegraphics[width=4.5em]{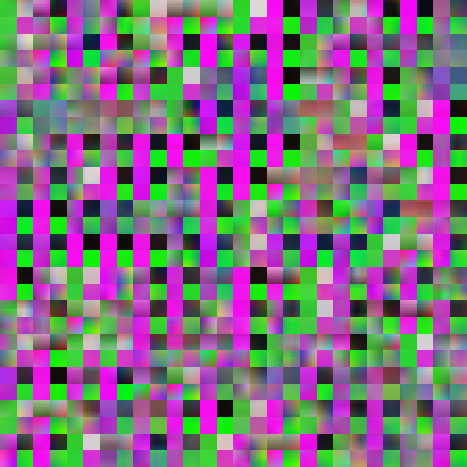}}
\newcommand{\cencsttwo}{\includegraphics[width=4.5em]{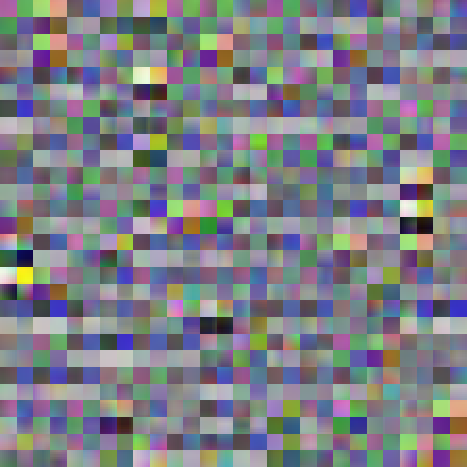}}
\newcommand{\cencstthree}{\includegraphics[width=4.5em]{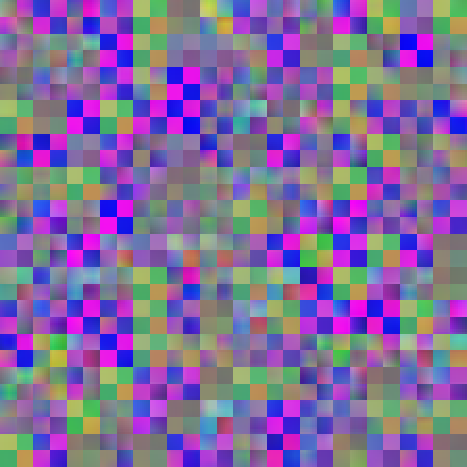}}
\newcommand{\cencstfour}{\includegraphics[width=4.5em]{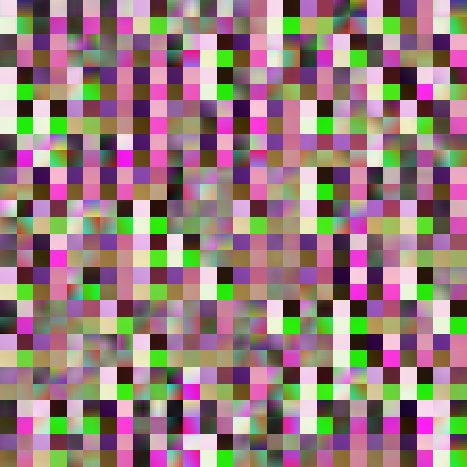}}

\newcommand{\cresonestone}{\includegraphics[width=4.5em]{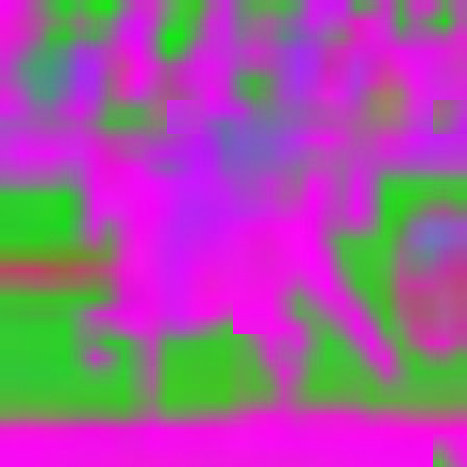}}
\newcommand{\cresonesttwo}{\includegraphics[width=4.5em]{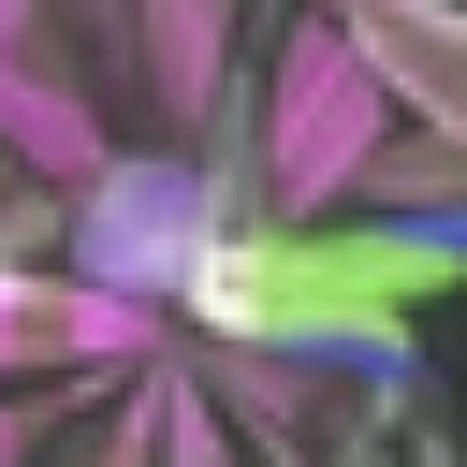}}
\newcommand{\cresonestthree}{\includegraphics[width=4.5em]{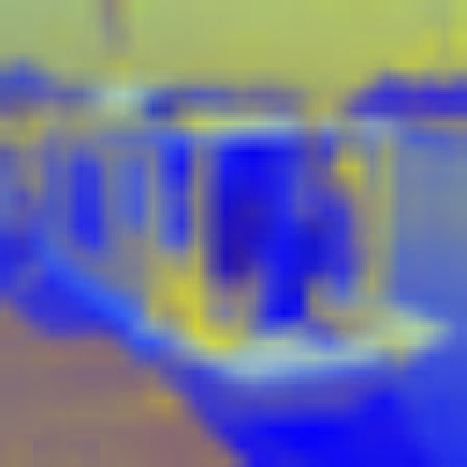}}
\newcommand{\cresonestfour}{\includegraphics[width=4.5em]{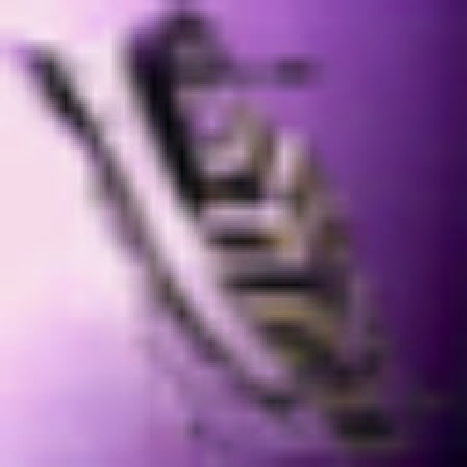}}

\newcommand{\crestwostone}{\includegraphics[width=4.5em]{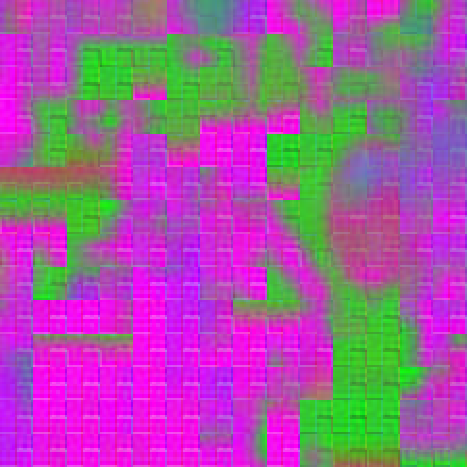}}
\newcommand{\crestwosttwo}{\includegraphics[width=4.5em]{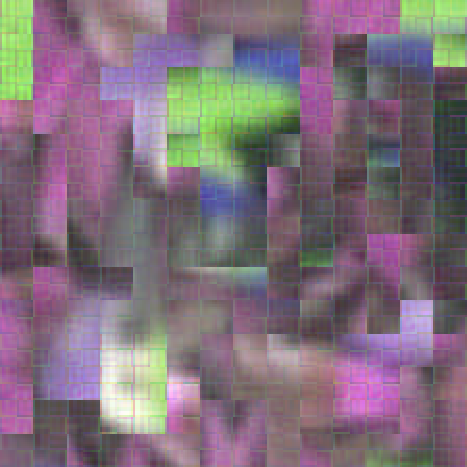}}
\newcommand{\crestwostthree}{\includegraphics[width=4.5em]{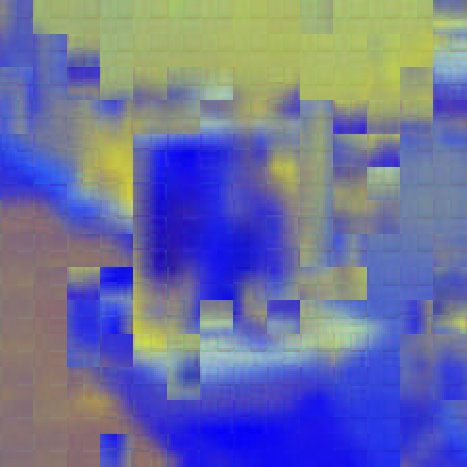}}
\newcommand{\crestwostfour}{\includegraphics[width=4.5em]{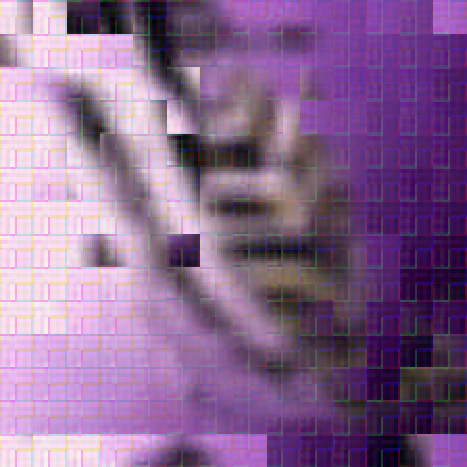}}

\newcolumntype{C}{>{\centering\arraybackslash}m{4.5em}}

\begin{figure*}[t]
	\centering
	\begin{tabular}{CCCCC|CCCC}
	Dataset & \multicolumn{4}{c}{CIFAR-10} & \multicolumn{4}{c}{STL-10}\\
	\toprule
	Original &\coristone & \coristtwo & \coristthree & \coristfour & \oristone& \oristtwo & \oristthree & \oristfour\\
	Encrypted &\cencstone & \cencsttwo & \cencstthree & \cencstfour & \encstone& \encsttwo & \encstthree & \encstfour\\ 
	Restored\par ($Q_{f}$=70, \par $S_{r}$=4:4:4) &\cresonestone & \cresonesttwo & \cresonestthree & \cresonestfour & \resonestone& \resonesttwo & \resonestthree & \resonestfour\\
	 & $1.00$ & $1.00$ & $0.99$ & $1.00$ & $1.00$ & $1.00$ & $0.92$ & $0.54$ \\
	Restored\par ($Q_{f}$=70, \par $S_{r}$=4:2:0) &\crestwostone & \crestwosttwo & \crestwostthree & \crestwostfour & \restwostone& \restwosttwo & \restwostthree & \restwostfour\\
	& $0.09$ & $0.23$ & $0.40$ & $0.46$ & $0.71$ & $0.63$ & $0.55$ & $0.01$ \\
	\bottomrule
	\end{tabular}
 \caption{Examples of images reconstructed from encrypted images by using jigsaw puzzle solver-based attack. Largest component ($L_{c}$) values are given under restored images}
	\label{fig:ex_result}
\end{figure*}

\section{Experiments and Results}
\subsection{Experimental Conditions}
In this section, the security of the compressible and learnable image encryption \cite{2023_Imaizumi} is discussed by using the jigsaw puzzle solver-based attack.
We evaluate the security of encrypted images with the following three metrics \cite{Cho_2010_CVPR,Gallagher_2012_CVPR}:
\\
{\bf Direct comparison ($D_{c}$):} represents the ratio of the number of blocks which are in the correct position.
\\
{\bf Neighbor comparison ($N_{c}$):} indicates the ratio of the number of correctly joined blocks.
\\
{\bf Largest component ($L_{c}$):} is the ratio of the number of the largest joined blocks that have correct adjacencies to the number of blocks in an image.
\par
In the measure, $D_{c}, N_{c}, L_{c} \in [0,1]$, a smaller value means the difficulty of recognizing objects.
We used 1000 images chosen from the CIFAR-10 and STL-10 datasets independently and each image was resized to 224$\times$224 pixels before encryption.
Each encrypted image was evaluated after JPEG-compression with the quality factor $Q_{f}=70,80,90$ and sub-sampling ratio $S_{r}=$ 4:4:4, 4:2:0 by using the IJG (Independent JPEG Group) software \cite{JPEGLIB}.
Ten different encrypted images were generated from one ordinary image by using different secret keys.


\subsection{Experimental Results}
Fig. \ref{fig:dclc} shows the result of security evaluation of the compressible and learnable image encryption against the jigsaw puzzle solver-based attack.
As illustrated in Fig. \ref{fig:dclc}, encrypted images were restored by using the jigsaw puzzle solver-based attack even when the low quality factor ($Q_{f}=70$) is used for JPEG-compression with 4:4:4 sub-sampling ratio.
Although it was confirmed that the use of 4:2:0 sub-sampling ratio enhances robustness than 4:4:4 sub-sampling ratio, some images were partially restored as $L_{c}=0.4$.
Fig. \ref{fig:ex_result} shows the examples of images reconstructed by using the jigsaw puzzle solver-based attack.
As shown in Fig. \ref{fig:ex_result}, it was confirmed that not only the CIFAR-10 dataset, but also the encrypted images from the STL-10 dataset can be restored by the attack.


\section{Conclusion}
In this paper, we evaluated the security of the compressible and learnable image encryption against jigsaw puzzle solver-based attack.
Experimental results showed that the use of the attack enables us to restore visual information on encrypted images under JPEG-compression with 4:4:4 sub-sampling ratio.
Although encrypted images compressed with 4:2:0 sub-sampling ratio enhance security, it was confirmed that some images were partially restored by using the attack.

\section*{Acknowledgment}
This study was partially supported by JSPS KAKENHI (Grant Number JP21H01327) and the Support Center for Advanced Telecommunications Technology Research, Foundation (SCAT).

\bibliographystyle{IEEEbib}
\bibliography{refs}
\end{document}